\documentclass[doublecol]{epl2} 

\usepackage[caption = false]{subfig}%
\usepackage{url}%
\usepackage{mathtools}
\usepackage{flushend}
\usepackage{mathtools, cuted}
\usepackage{stfloats}
\usepackage{float}
\title{Frictional Damping in Biomimetic Scale Beam Oscillations}
\shorttitle{Frictional Damping in Biomimetic Scale Beam Oscillations} 

\author{Hessein Ali \inst{1} \and Hossein Ebrahimi\inst{1} \and Ranajay Ghosh\inst{1,a}}
\shortauthor{H. Ali \etal}

\institute{                    
  \inst{1} Department of Mechanical and Aerospace Engineering, University of Central Florida, Orlando FL \\
  \inst{a} Email:ranajay.ghosh@ucf.edu
}

\abstract{
Stiff scales adorn the exterior surfaces of fishes, snakes, and many reptiles. They provide protection from external piercing attacks and control over global deformation behavior to aid locomotion, slithering, and swimming across a wide range of environmental condition. In this letter, we investigate the dynamic behavior of biomimetic scale substrates for further understanding the origins of the nonlinearity that involve various aspect of scales interaction, sliding kinematics, interfacial friction, and their combination. Particularly, we study the vibrational characteristics through an analytical model and numerical investigations for the case of a simply supported scale covered beam. Our results reveal for the first time that biomimetic scale beams exhibit viscous damping behavior even when only Coulomb friction is postulated for free vibrations. We anticipate and quantify the anisotropy in the  damping behavior with respect to curvature. We also find that unlike static pure bending where friction increases bending stiffness, a corresponding increase in natural frequency for the dynamic case does not arise for simply supported beam. Since both scale geometry, distribution and interfacial properties can be easily tailored, our study indicates a biomimetic strategy to design exceptional synthetic materials with tailorable damping behavior.}

\begin{document}

\maketitle
Dermal scales are one of the oldest evolutionary adaptation in animals and found in many organisms of the kingdom Animalia including fishes~\cite{C1,C2}, reptiles ~\cite{C3,C4} and mammals~\cite{C5,C6}. Due to their ubiquity and sheer longevity in evolutionary history, scales have been garnering intense scrutiny from materials and structures community~\cite{C7,C8,C9}. From a materials perspective, characterization studies of various scales have revealed highly intricate microstructures. Such scales can be hybrid ~\cite{C10,C11,C12}, hierarchical~\cite{C13,C14,C15}, and composite in nature ~\cite{C16,C17,C18} capable of engaging multiple length scales depending on the load~\cite{C19,C20,C21}. Scales are also naturally multifunctional~\cite{C22,C23,C24,C25}. The most apparent function is the protective properties of scales for the underlying substrate, which has been an inspiration of ancient armor designs~\cite{C26}.  In addition to protection against external objects, which is mechanically an indentation type local deformation problem, global deformation modes such as bending and twisting of a substrate reveal equally interesting properties. These include reversible nonlinear stiffening and locking behavior due to the sliding kinematics of the scales in one-dimensional substrates~\cite{C27,C28,C29,C30,C31,C32,C33}.  In spite of rapidly accumulating literature on the mechanics of scales and scale covered systems, their influence on the dynamic behavior of substrates has not been revealed beyond characterizing strain rate effects under transient loading ~\cite{C34,C35,C36}. 

In sustained dynamic behavior such as oscillations, the role of scale interfacial friction can be critical in affecting the long-term response. This type of behavior is especially apparent for applications such as swimming, locomotion and structural vibration. Recent research for the static pure bending case showed that dry friction adds resistance to scale sliding inevitably increases bending stiffness~\cite{C28}. Thus there seems to be an inherent interplay between the stiffness, geometry, range of motion and friction behavior in these type of systems.

In this letter, we study for the first time the damped oscillations of biomimetic scale beam  using a combination of numerical and analytical modeling. The work addresses the first mode of bending vibrations.

\begin{figure*}[t]
\centering
\subfloat[]{%
\includegraphics[scale = 0.46]{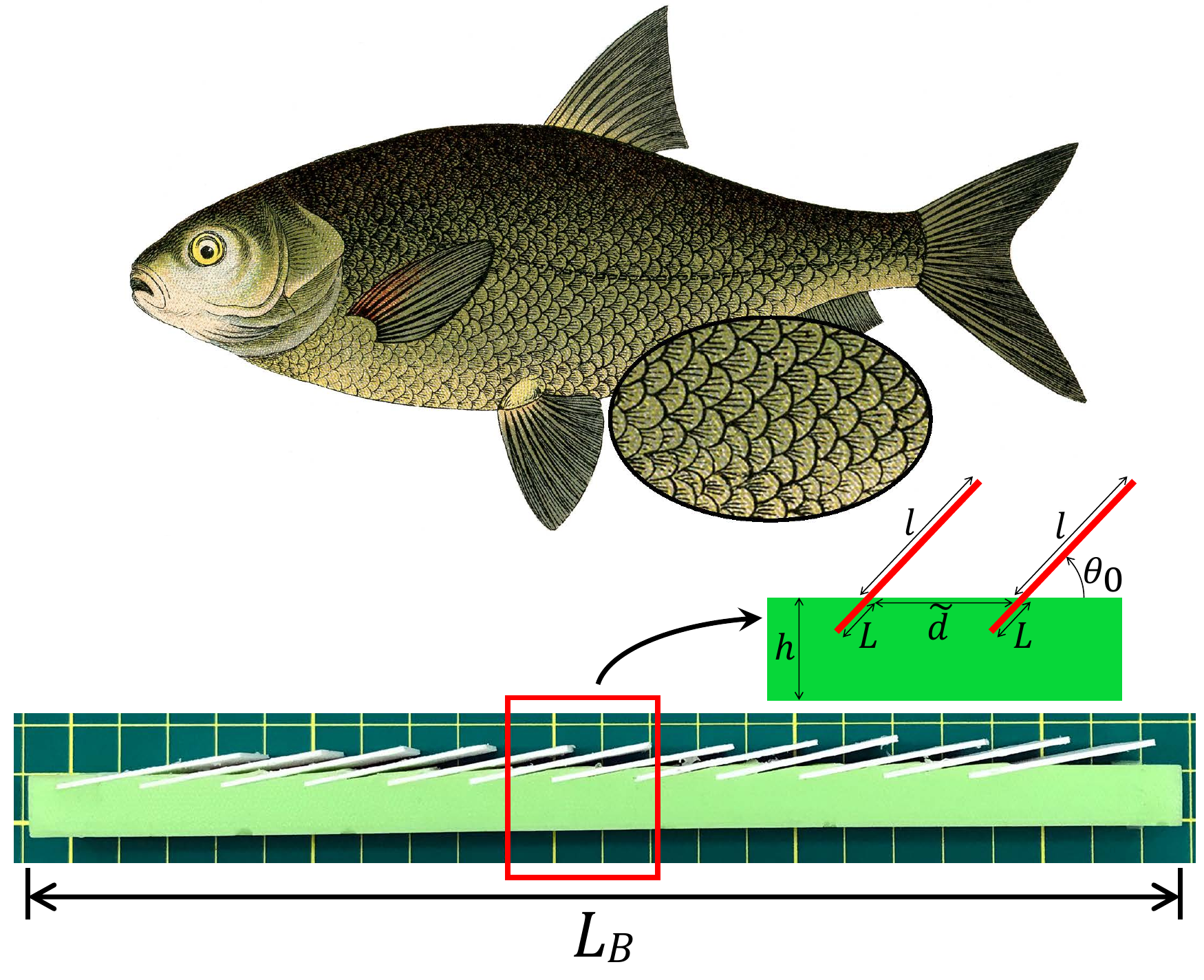}
\label{F1a}}
\quad
\hspace{-20pt}
\subfloat[]{%
\includegraphics[scale=0.46]{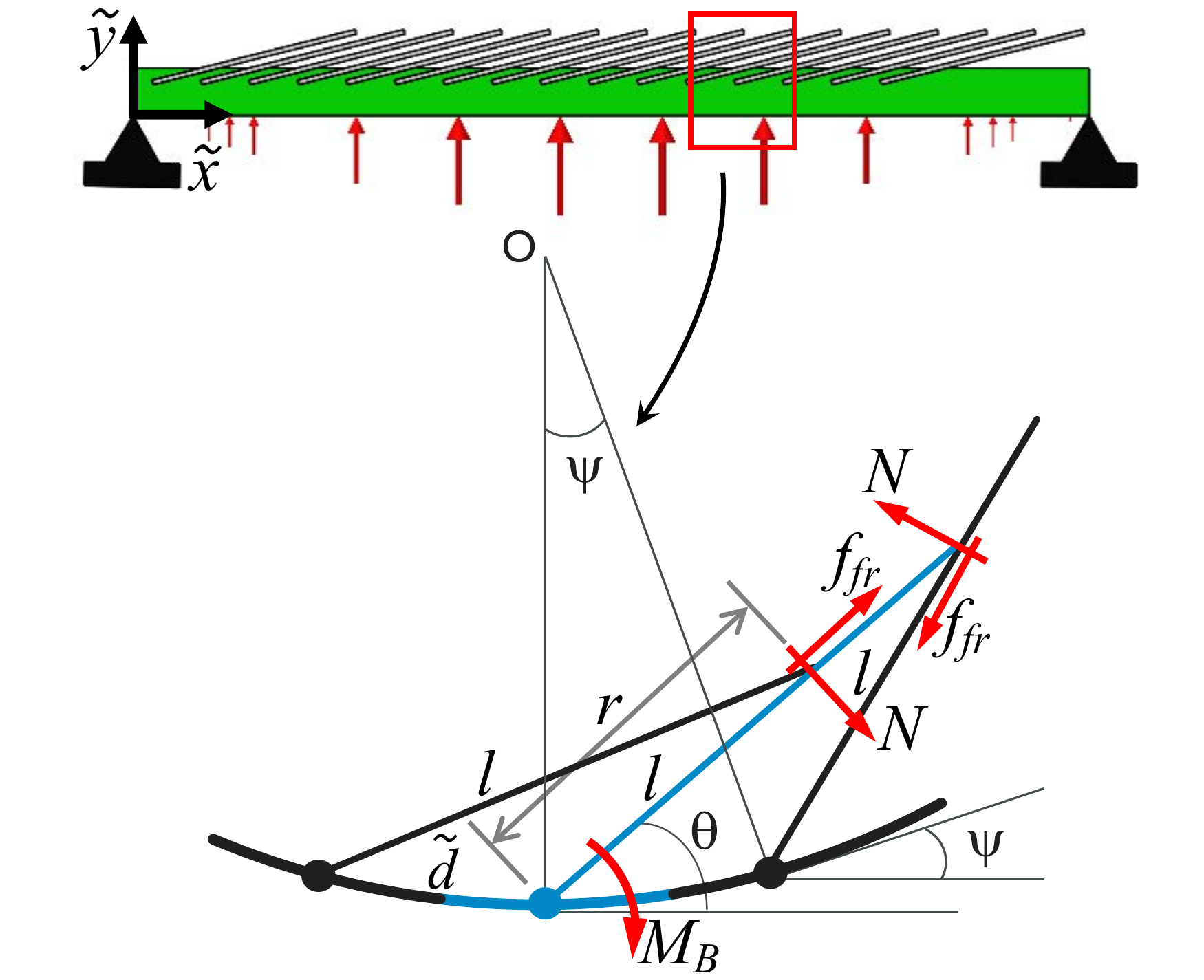}
\label{F1b}}
\caption{(a) Natural history fish image illustrating a typical scale covered skin~\cite{C0} and a biomimetic scale covered beam. (b) Schematic diagram of simply supported biomimetic scale covered beam and RVE geometry.}
\vspace{-10pt} 
\label{fig1} 
\end{figure*}

We first simplify the complex structure of biomimetic scale beam by assuming a linear-elastic substrate with partially embedded rigid scales on its top surface Fig.~\ref{F1a}. This is valid because the scales used in this study are several orders of magnitude stiffer than the substrate material~\cite{C27,C29} and typical substrate strains are moderate, with excessive deformations arrested by additional stiffness induced by the stiff scales. This assumption also leads to isolation of nonlinearity stemming purely from scales interaction, rather than material effects. We assume that the substrate is of length $L_B$ and height $h$ and with Young's modulus $E_B=1.5$ MPa, Poisson's ratio $\nu=0.42$~\cite{C31,C33}, and density $\rho_B=854 kg/m^3$. Although these values are not critical to the outcomes of the study, these are typical of silicone polymer (Vinylpolysiloxane) used for fabricating many soft substrates. Each individual scale is of thickness $D$, density $\rho_s$ , and total length $l_s$ that consists of an exposed part $l$ and embedded part $L$, or $l_s=l+L$, Fig.~\ref{F1a}. In addition, we impose $D \ll l_s$ and $L \ll h$, (shallow embedding). This allows each individual scale to be modeled as a linear torsional spring rotating about fixed point with constant $\tilde{K}_s$ ~\cite{C27,C29,C31}. This constant was analytically computed as ${K}_s=E_B D^2 C_B (L/D)^n$ where $C_B$,$n$ are constants with corresponding values $0.66$ and $1.75$, respectively for same materials earlier~\cite{C27}. The ratio of scale length to spacing is denoted  by $\eta=l/\tilde{d}$,  where $\tilde{d}$ is the distance between any two neighboring scales~\cite{C27}. The scales are embedded onto the substrate at an initial angle $\theta_0$, measured with respect to the beam's centerline, and increase nonlinearly once the engagement of scales commences ~\cite{C27,C29,C31}. Due to small strains in the beam substrate, we use Euler-Bernoulli beam assumptions.  
 
We also consider uniform scale distribution over the substrate. In this configuration, scales preserve periodicity post-engagement under pure bending. However, beyond this limited loading regime, such periodicity would not hold ~\cite{C31}. This prevents us from using the closed form of kinematic relationship existing in current literature~\cite{C27,C28,C29}. However, in the limit of sufficiently dense distribution of scales we can reasonably impose local periodicity, Fig.~\ref{F1b}. This assumption leads to a locally uniform pure bending mode thereby allowing previously developed kinematic relationships to be valid at least in a local sense. The global deformation of the scaly beam can be envisioned as a combination of substrate bending and scales rotation~\cite{C27,C31}. Accordingly, the energy stored in the system is an additive combination of the two deformation modes. Friction is modeled using Coulomb dry friction with a coefficient of friction $\mu$.

The local periodicity assumption introduced earlier allows the isolation of a representative volume element (RVE), Fig.~\ref{F1b}. Thus, locally the kinematic relationship can be given by $\theta_{RVE}= \sin^{-1}{(\eta \psi_{RVE} \cos{\psi_{RVE}/2})-\psi_{RVE}/2}$ ~\cite{C28} where $\psi_{RVE}= \tilde{\kappa}_{RVE}\tilde{d}$ is the angular rotation of each RVE measured with respect to the beam's instantaneous curvature $\tilde{\kappa}_{RVE}$, Fig.~\ref{F1b}. 

We derive the dynamic equation of motion by invoking the Hamilton's principle of least action leading to the variational energy equation   $\delta \int_{\tilde{t}_1}^{\tilde{t}_2}{(\hat{T}-\hat{V}+W})d\tilde{t} = 0$ for the entire beam.  Here $\hat{T}$ is the total kinetic energy per unit length,$\hat{V}$ is the total strain energy per unit length, while the term $W$ denotes the work done by the applied traction (see supplemental material (SM) for their formulas). We apply a harmonic load on the smooth (side opposite to scales) of form $\tilde{f}(\tilde{x},\tilde{t}) = \tilde{f}_0 \phi(\tilde{x})\cos{\tilde{\Omega}\tilde{t}}$ to ensure that only the first mode of vibration is activated~\cite{C39} with simply supported boundary conditions. Here $\tilde{f}_0$ is the load amplitude, $\phi(\tilde{x})$ is the shape function of the first eigenmode of simply supported beam ~\cite{C40}, while $\tilde{\Omega}$ is the load frequency. This work neglects the mass of scales (See SM for justification). 

\begin{figure*}[h]
\begin{equation}
\label{Eq1}
\begin{aligned}
  & {\rho _B}A{{{\partial ^2}\tilde y} \over {\partial {{\tilde t}^2}}}\; + {E_B}{I_B}{{{\partial ^4}\tilde y} \over {\partial {{\tilde x}^4}}}\; + \tilde C{{\partial \tilde y} \over {\partial \tilde t}} + {{{\partial ^2}} \over {\partial {{\tilde x}^2}}}{1 \over {{N_{RVE}}}}\bigg{[}\mathop \sum \limits_{RVE} {{\tilde K}_s}\left( {{\theta _{RVE}} - {\theta _0}} \right) {{\partial {\theta _{RVE}}} \over {\partial {\psi _{RVE}}}} + \cr
  & \mathop \sum \limits_{RVE} {{\sin \left( \beta  \right){{\tilde K}_s}\left( {{\theta _{RVE}} - {\theta _0}} \right)\;{\rm{sgn}}\left( {\mathop {\dot{\tilde y}}\limits^{} } \right)} \over {\cos \left( {{\psi _{RVE}} + \beta } \right) - \bar r\cos \left( \beta  \right)}} \bigg{]} H\left( {{{\tilde \kappa }_{RVE}} - {{\tilde \kappa }_e}} \right)   
   = f\left( {\tilde x,\tilde t} \right). 
\vspace{-10pt}
\end{aligned}
\end{equation}
\end{figure*}

The Hamilton’s principle leads to the differential equation expressed in Eq.\ref{Eq1} (See SM for derivation). Where $A$ refers to the cross sectional area of the substrate, $I_B$ is the second moment of area, $\beta=\tan^{-1}{\mu}$ is the scales friction angle, and $\bar{r}$ is the distance per scales length measured from scales base to the interaction with neighboring scale (Fig.~\ref{F1b}). The Heaviside step function ensures that scales contribution is accounted for only in the case when the beam is deflected downward and when $\theta_{RVE} \ge \theta_0$. Note that $\tilde{\kappa}_e$ refers to the RVE curvature at the instant $\theta_{RVE} = \theta_0$. A small artificial viscous damping term with constant $\tilde{C}$ is included in the equation to model any other source of damping and to stabilize numerical integrators for the forced oscillation cases. The variable $N_{RVE}$ refers to the number of RVEs utilized in the solution of the system. Note that the results would not depend on the actual number of RVEs for a sufficiently large number of RVEs. In our case, we found $N_{RVE}=10$ to be sufficient. In this equation, the terms inside the summations agree with the quasi-static moment equations expressions derived earlier in literature~\cite{C27,C28}.

The governing Equation~\ref{Eq1} can be written in a non-dimensional form by first noting that two natural length scales emerge from the geometry, the length of beam $L_B$ (horizontal length scales) and the radius of gyration of the cross section $R=\sqrt{I_B/A}$ (vertical length scales). We denote the ratio of the two as $\gamma=R/L_B$. In addition, a natural time scale for the problem can be extracted from the natural frequency expression of the underlying elastic substrate, $\zeta=\sqrt{\rho_B AL_B^4/E_B I_B}$.  This allows us to non-dimensionalize the following quantities appearing in Eq.~\ref{Eq1} $x=\tilde{x}/L_B$, $d=\tilde{d}/L_B$, $y=\tilde{y}/R$, $t=\tilde{t}/\zeta$ and $\Omega=\tilde{\Omega}\zeta$. These intrinsic length and time scales can also be used to obtain the vertical force normalization unit by noting that Force $\sim MLT^{-2}=\rho_B AL_B R\zeta^{-2}$. This leads to a vertical traction normalizer given by $F'=E_B I_B R/L_B^4$ and thus $f_0=\tilde{f}_0/{F'}$. Similar arguments also lead to normalization of other variables $C={\tilde{C} L_B^2 \over \sqrt{\rho_B AE_B I_B }}$ and $K_s={\tilde{K}_s L_B \over E_B I_B}$.  With these non-dimensionalizations, the non-dimensionalized equation of motion (EOM) is now

\begin{gather}
   {{{\partial ^2}y} \over {\partial {t^2}}}\; + {{{\partial ^4}y} \over {\partial {x^4}}}\; + C{{\partial y} \over {\partial t}} + {{{\partial ^2}} \over {\partial \notag {x^2}\;}}{1 \over {\gamma {N_{RVE}}}}\bigg{[}\mathop \sum \limits_{RVE} {K_s}({{\theta _{RVE}}  \notag
   - {\theta _0}})\notag \\
   {{\partial {\theta _{RVE}}} \over {\partial {\psi _{RVE}}}} +  \notag
   \mathop \sum \limits_{RVE} {{\sin \left( \beta  \right){K_s}\left( {{\theta _{RVE}} - {\theta _0}} \right)\;{\rm{sgn}}\left( {\mathop {\dot{\tilde y}}\limits^ {}}  \right)} \over {\cos \left( {{\psi _{RVE}} + \beta } \right) - \bar r\cos \left( \beta  \right)}}\bigg{]}  \\
    H\left( {{\kappa _{RVE}} - {\kappa _e}} \right) 
   = f\left( {x,t} \right). 
	\label{Eq2}
	\end{gather}

For simply supported beam, the boundary conditions are $y=0$ and $\partial^2 y / \partial x^2 =0$ at $x=0$ and $1$. Two types of initial conditions would be studied, displacement and velocity. The displacement type would be denoted by $y(x,0)=AA(\phi(x))$ where $AA$ is a constant and $\phi(x)$ is the non-dimensional counterpart of the shape function introduced earlier. The velocity initial condition is given by $v(x,0)=BB(\phi(x))$ where $BB$ is a constant.  Equation~\ref{Eq2} can be solved by utilizing separation of variables $y(x,t)=\phi(x)T(t)$. Here $T(t)$ is a time-dependent function while $\phi(x)$ is the first eignmode of a simply supported beam, which is known to be $\sin{\pi x}$ ~\cite{C40}. Using Bubnov-Galerkin weighted residual method we obtain the following second order nonlinear ordinary differential equation for the time-dependent part (See Supplemental Information for detailed derivation)

\begin{equation}
\begin{aligned}
  & \ddot T + C\dot T + {\pi ^4}T + {K_s}{1 \over {{N_{RVE}}}}\bigg{[}\mathop \sum \limits_{RVE} {f_1}\left( {T,\eta ,d,{T_e},\gamma } \right) + \cr 
  &  \mathop \sum \limits_{RVE} {f_2}\left( {T,\eta ,d,{T_e},\gamma ,\mu } \right)\;{\rm{sgn}}\left( {{{\dot T}_{RVE}}} \right)\bigg{]}H\left( {{\kappa _{RVE}} - {\kappa _e}} \right)  \cr
 & =  {f_0}\cos {\rm{\Omega }}t 
	\label{Eq3}
\end{aligned}
\end{equation}

Here, $f_1$ and $f_2$ are complex nonlinear functions (see SM for their form). This nonlinear ordinary differential equation is solved using a direct numerical integrator such as the Newmark beta scheme which nests a Newton-Raphson nonlinear equation solver~\cite{C41}.

The results of our mathematical model are verified numerically via FE simulations using a commercially available FE software ABAQUS (Dassault systems). In the FE model, an assembly of 2D deformable shell parts (for scales and substrate) was made with rigidity imposed on the embedded scales and linear-elastic properties were assigned for the substrate. An implicit dynamic step was then utilized. The initial conditions were imposed through a predefined velocity in the form $\dot{T}(0)=BB(\phi(x))$ where BB is the magnitude of velocity. The imposition of contact between scales, mesh convergence, and boundary condition were done consistent with literature~\cite{C31}. Interfacial friction was introduced using penalty function formulation with coefficient $\mu$~\cite{C42}.

\begin{figure}[t]
\centering
\subfloat[]{%
\includegraphics[scale = 0.5]{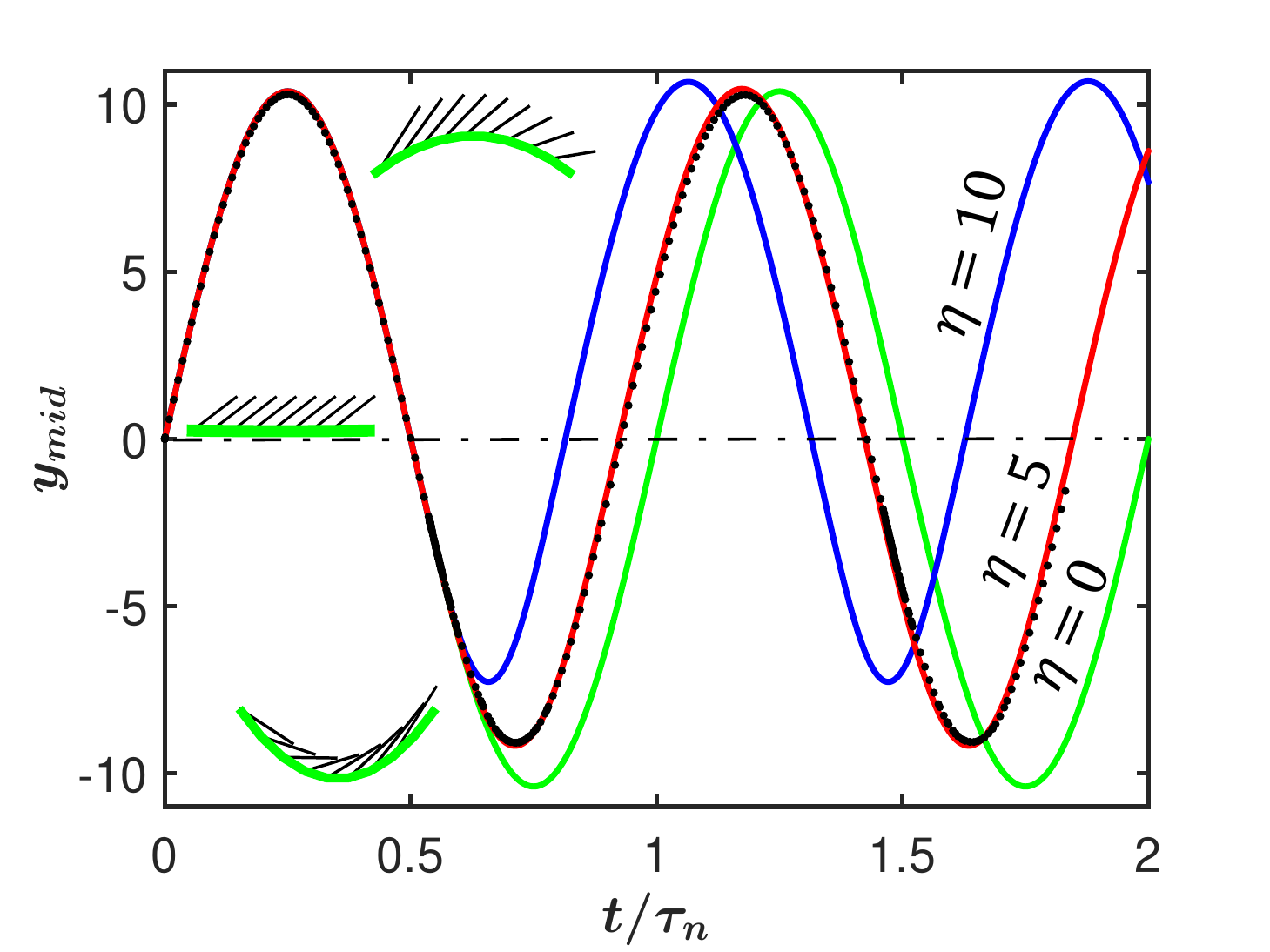}
\label{F2a}}
\quad
\hspace{-20pt}
\subfloat[]{%
\includegraphics[scale=0.5]{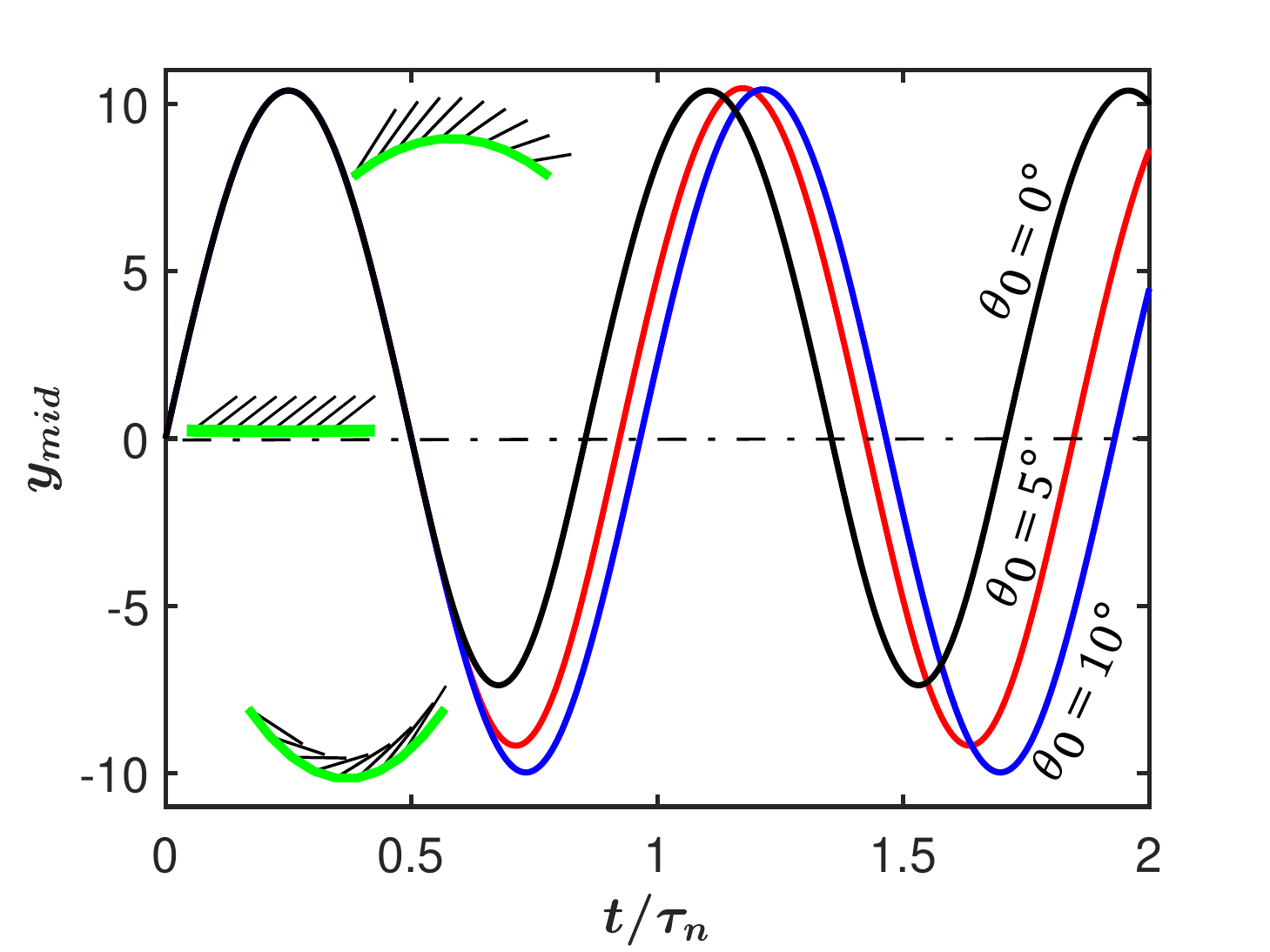}
\label{F2b}}
\caption{(a) The response of biomimetic scale beam under velocity initial condition for $\theta_0 = 5^\circ$ and various $\eta$ (The block dots represent FE simulations). (b) The effect of $\theta_0$ on the behavior of biomimetic scale beam with $\eta=5$ and under velocity initial condition.  For these plots and all others, the number of RVE was $10$ in the model. Insets show the concave, neutral, and convex side of deformation.}
\vspace{-10pt} 
\label{F2} 
\end{figure}

In the small deformation regime, two important geometrical parameters are of interest in these simulations, $\eta$ and initial angle $ \theta_0$. We compare and contrast the effect of these two sources in free vibration regime $f_0=0$ in Eq.~\ref{Eq3} in Figs.~\ref{F2a} and \ref{F2b}. In these simulations, an initial velocity $BB=10.4$ (chosen to assure engagement of scales for the assigned $\theta_0$) is applied to the beam and the resulting normalized mid-point deflections are plotted with normalized time. Here the time is normalized by the undamped period of a plain beam $\tau_n=2\pi/w_n$ , $w_n=\pi^2$. We do not use any artificial viscous damping for free vibration simulations. 

We first investigate the effect of geometric parameters on the free vibration of the biomimetic beam without any friction. We denote the side with scale engagements as the concave side and the one where scales do not engage as convex, see Fig.~\ref{F2a} inset. In Fig.~\ref{F2a}, we plot the midpoint deflection of the biomimetic beam without taking the effect of friction with increasing overlap ratio $\eta$. There are two visible effects of increasing $\eta$ on the free vibration regime. The first is the expected increase in the natural frequency of vibration due to the added stiffness. However, increasing $\eta$ also decreases the amplitudes of vibration on the concave side, with little effect in the deflection on the convex side. We expect this behavior since higher $\eta$ leads to greater energy stored in scales rotation compared to the substrate. Note that the change in $\eta$ corresponds to varying d and fixing $l$ for example $\eta=5$ corresponds to $d=0.05$ and $\eta=0$ refers to a plain beam. Our results are an excellent match with FE simulations, which are depicted by black dots for $\eta=5$. Therefore, overall, higher overlap ratio leads to a higher frequency and lower mean amplitude of vibration for same initial velocity profile. The other geometrical parameter of consequence is the initial inclination angle $\theta_0$. We fix the overlap ratio to $\eta=5$ and simulate the free vibration response under same initial conditions as before.  As inclination angle increases, the scales engagement is delayed. Therefore, the overall stiffness of the system is lesser at a given displacement. This leads to decrease in frequency of response as seen in Fig.~\ref{F2b}. The plots also reveal that the asymmetry in deflection between the scale side and plain side is made less severe with an increase in scale angle. For the case with $\theta_0=10^\circ$, no engagement takes place for the given initial condition and the vibrational symmetry is fully restored. In summary, these simulations highlight the critical role played by geometrical parameters of the scale $\eta$ and $\theta_0$ in tailoring the frequency and symmetry of vibrating biomimetic beam, analogous to deformation in the static regime~\cite{C27,C31}. 

\begin{figure}[htbp]
\centering
\subfloat[]{%
\includegraphics[scale = 0.5]{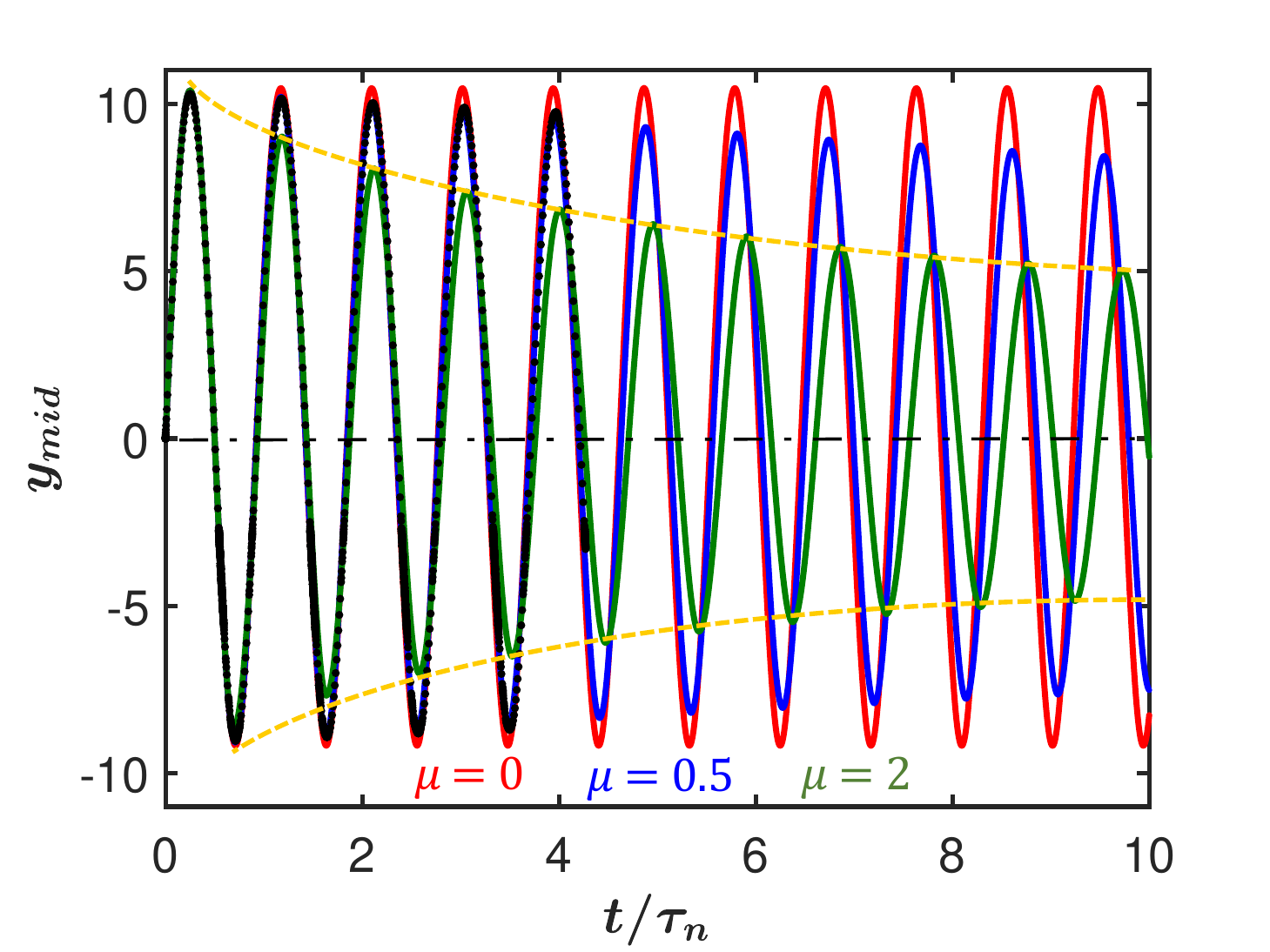}
\label{F3a}}
\quad
\hspace{-20pt}
\subfloat[]{%
\includegraphics[scale=0.5]{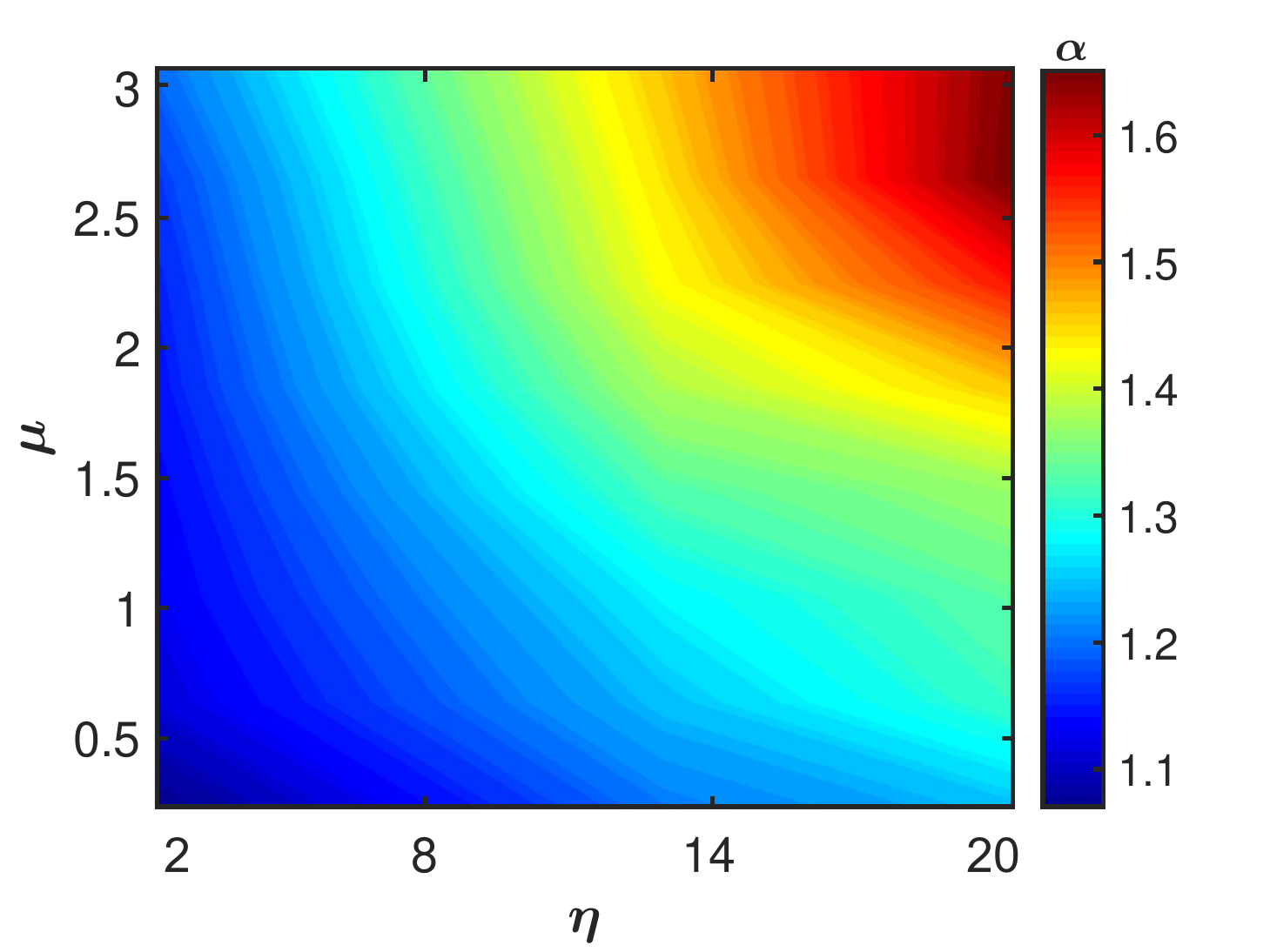}
\label{F3b}}
\caption{(a) The response of the scaly beam with $\eta= 5$ under velocity initial condition for various $\mu$ (The block dots represent FE simulations). (b) phase map of the ratio between the convex and concave damping coefficient ($\alpha$) spanned by $\eta$ and $\mu$. For these plots and all others, the number of RVE was $10$ in the model.}
\vspace{-10pt} 
\label{fig3} 
\end{figure}

We now study the interplay of friction with the critical geometric parameters for free vibrations under same initial conditions. In Fig.~\ref{F3a}, we fix $\eta=5$ , $\theta_0=5^\circ$ and vary the coefficient of friction. The black dots for $\eta=5$ in the figure indicate FE simulations, which are in excellent agreement with our model results. These results highlight several interesting phenomena. First, as expected we find a pronounced damping behavior with increasing $\mu$. However, unlike a dry friction spring-mass damper system, here the amplitude decay is not linear but exponential resembling a viscous damping behavior. This indicates a geometrically induced regime transformation of damping from dry friction to viscous drag. We also observe a clear increase in the time-period of oscillations with increasing $\mu$ as expected in viscous damped system. The similarities with conventional viscously damped system ends here. In typical conventional damped oscillators, the decaying behavior continues till motion is altogether arrested. In contrast, for the biomimetic system, the decay would not end in a complete stop but rather till deflection is small enough to preclude scales engagement, returning it to the frictionless conservative system, vibrating indefinitely.  Increasing friction also decreases the amplitudes of vibration, mirroring the stiffening effect of scales overlap observed in Fig.~\ref{F3a}, but here, the amplitude reduction occurs on both sides (concave and convex) of beam and not just on one side. This apparent distinction between the two stiffening mechanisms has to do with continuous energy loss brought about in every cycle in the damped oscillation case, lowering the overall amplitude in contrast to the frictionless case explored earlier. Furthermore, for this initial velocity oscillation, the natural frequency decreases with increasnig friction. Interesting this is in contrast to the stiffening role of friction found in the static case~\cite{C28}. Thus, in the dynamic case, no additional stiffening effect is apparent. 
In spite of friction affecting the amplitude of both sides of vibration, there is still an observed asymmetry with more rapid reduction occurring on the convex side. To investigate this asymmetry, we use a logarithmic decrement defined as $\delta={1 \over \Delta}\ln(AA_{n+1}/AA_n )$ where $\Delta$ is the time between $n$ and $n+1$ peaks for both sides of oscillations. We then take a ratio of the two to quantify the asymmetry, with $\alpha =\delta_{convex}/\delta_{concave}$. This quantity is used to generate an asymmetry phase map, Fig.~\ref{F3b} which is mapped by $\mu$ and $\eta$. This phase diagram indicates that anisotropy is considerably accentuated by increasing both $\eta$ and $\mu$. Specifically, the decays are larger on the convex side. Also, increasing the friction at lower $\eta$ hardly contributes to the asymmetry, presumably due to late stage scale engagement. On the other hand, higher $\eta$ amplifies even small friction coefficients appreciably. Overall, this study indicates that both distribution of the scales and its surface properties can be engineered to control the nature of oscillation of these systems. Another distinguishing feature of the system is the long-term behavior of damped oscillators. In classical damped systems, the oscillators eventually cease motion. However, in this case, a damped system first reduces the amplitude of motion. However, when the displacement is small enough to prevent further scales engagement, friction ceases to act and the system continues to oscillate linearly, indefinitely in the ideal case.

\begin{figure}[!h]
\centering
\subfloat[]{%
\includegraphics[scale = 0.5]{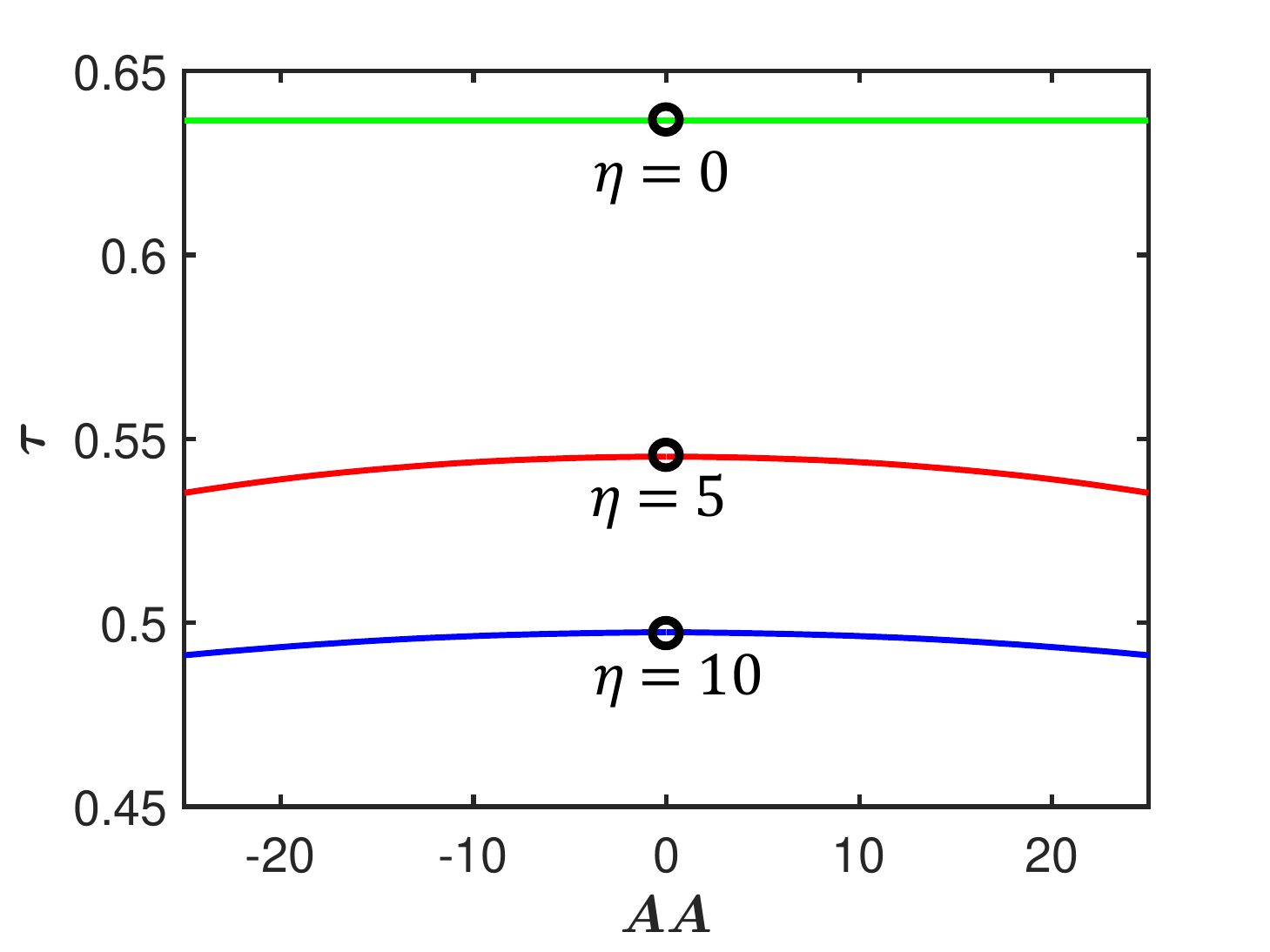}
\label{F4a}}
\quad
\hspace{-20pt}
\subfloat[]{%
\includegraphics[scale=0.5]{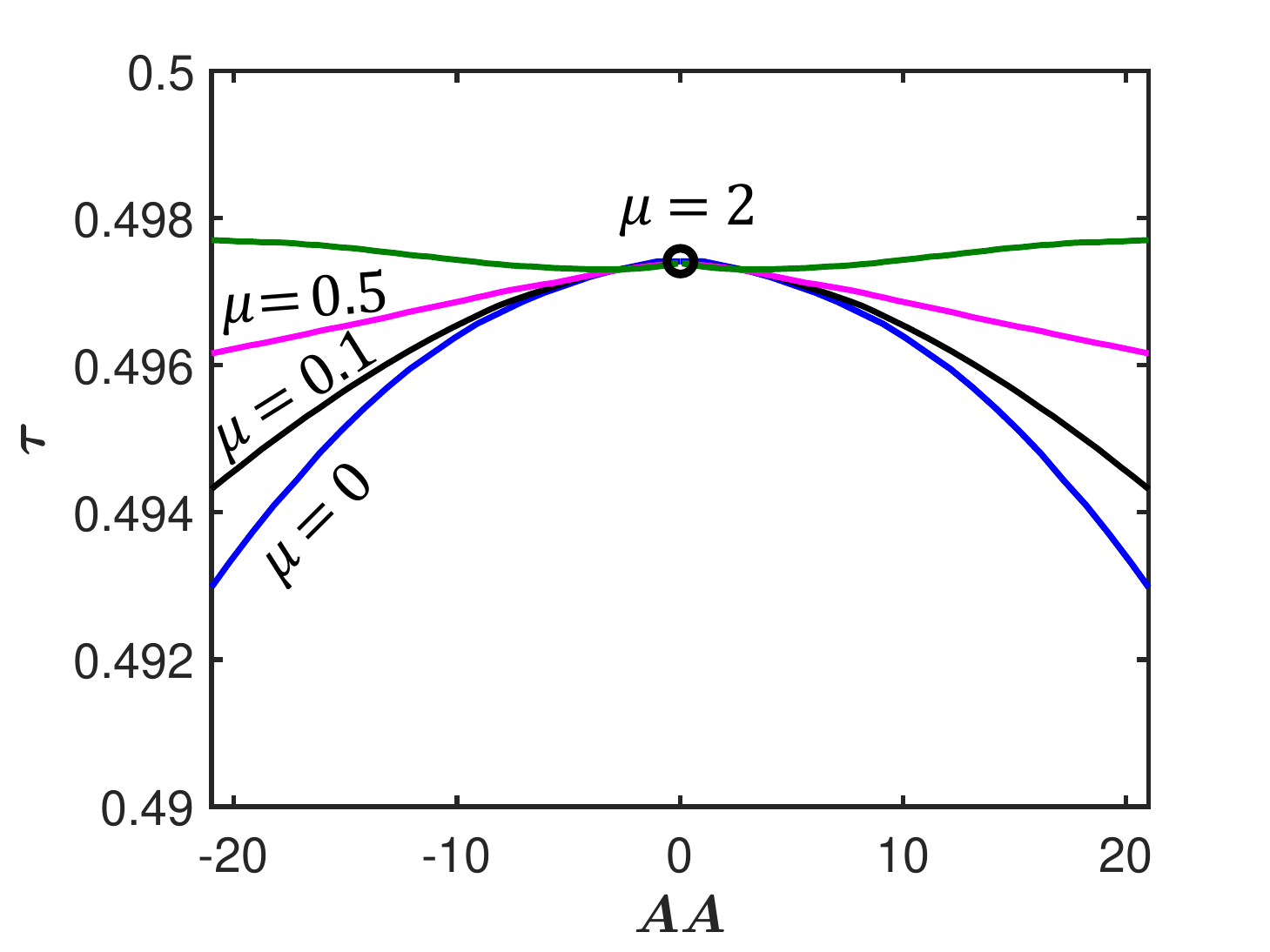}
\label{F4b}}
\caption{(a)  Period  of  free vibration  of  frictionless  biomimetic scale  beam  for various initial displacements $AA$ and $\eta$.  (b) Period  of  free vibration  of  frictional  biomimetic scale  beam  for various initial displacements $AA$ and $\mu$.  $\theta_0=0^\circ$ for both these simulations.}
\vspace{-10pt} 
\label{fig2} 
\end{figure}

Next, we turn our attention to the role of initial conditions in influencing the time-period of oscillations. We now give an initial displacement condition to the beam which is varied and corresponding time-period of oscillation calculated. We find that this variation leads to different time-periods for the system. We undertake two different studies to investigate the role of $\eta$ and $\mu$ on this phenomena. In Figure.\ref{F4a},  frictionless biomimetic scale beams are studied for various values of $\eta$. For a plain beam ($\eta$=0), there is a complete insensitivity of time-period on initial displacement, as expected. Note that we have ignored nonlinearity from large deformation or material sources in this case. As scales with higher $\eta$ are introduced, for a given initial displacement, the time period of oscillations are smaller for higher $\eta$. At the same time, we also observe that as $\eta$ increases, there is dramatic increase in time-period sensitivity to initial amplitude for a given beam, leading to pronounced decrease of time-period with increasing initial amplitude. However, this trend is gradually arrested as $\eta$ increases further since the system now begins to oscillate near its locking limit~\cite{C28}. This leads to restoration of insensitivity. Furthermore, this insensitivity restoration phenomena can also be introduced by increasing the coefficient of friction for a given $\eta$. This can be seen in Fig.~\ref{F4b} where we plot the dynamic response of a biomimetic beam with $\eta=10$ for different $\mu$. This trend continues till $\mu$ is large enough to transition the system into a really small deformation regime oscillation due to proximity to frictional locking curvature ($AA>20$ and $\mu=2$ ). Due to very high friction, the system quickly attenuates, returning to a linear stage due to lack of scale engagements. This results in almost no sensitivity to initial displacement. Overall, for a given initial condition, scales with higher $\mu$ have higher time periods. This is consistent with the results obtained for initial velocity simulations, ~\ref{F3b}. 

At this point it is interesting to compare this behavior with that of hardening Duffing oscillator which also exhibits this characteristic amplitude dependent time period~\cite{C43}. Although not exactly a Duffing system, we have in effect obtained a tailorable hardening spring system. 

\begin{figure}[htbp]
\centering
\subfloat[]{%
\includegraphics[scale = 0.5]{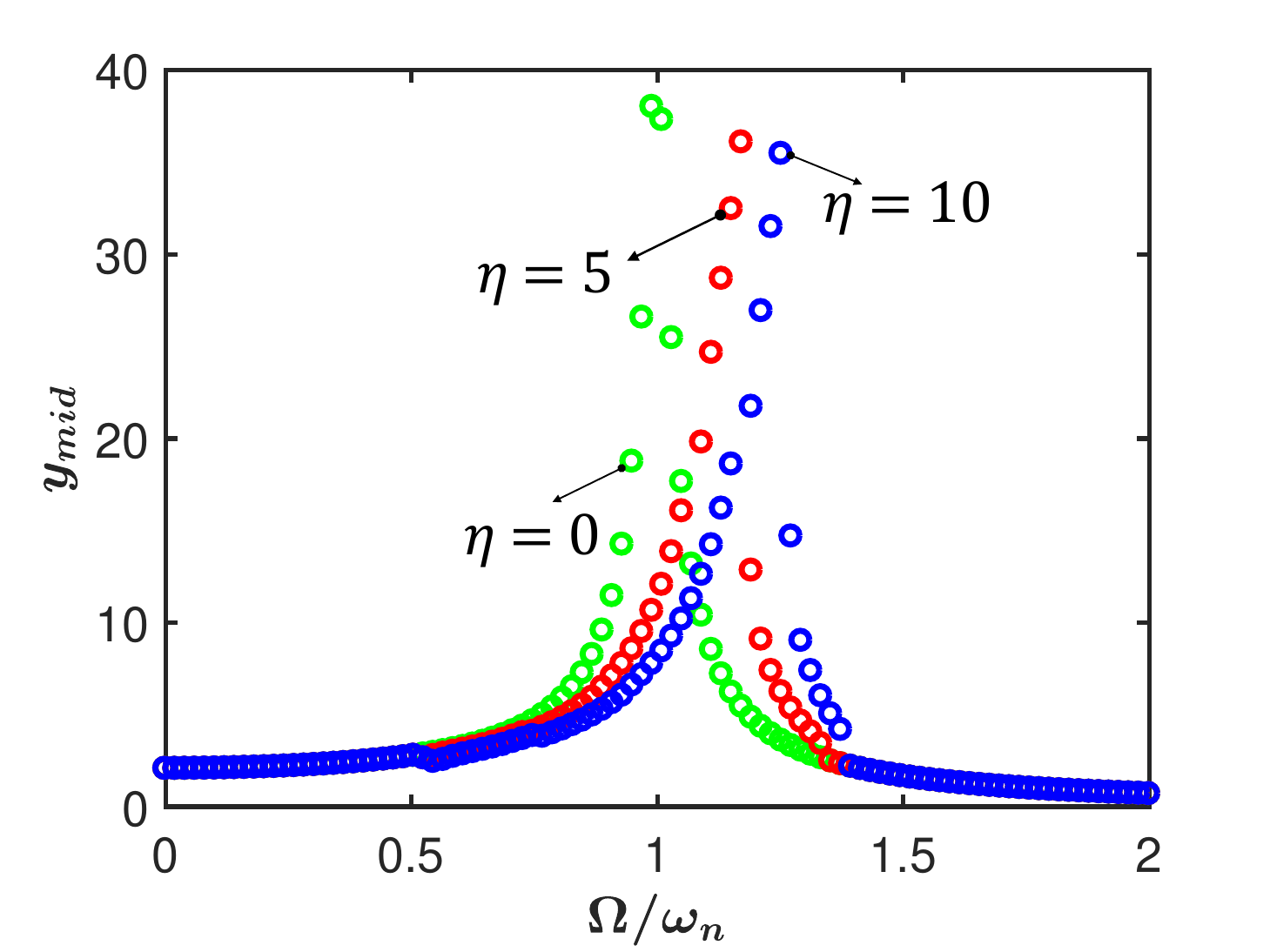}
\label{F5a}}
\quad
\hspace{-20pt}
\subfloat[]{%
\includegraphics[scale=0.5]{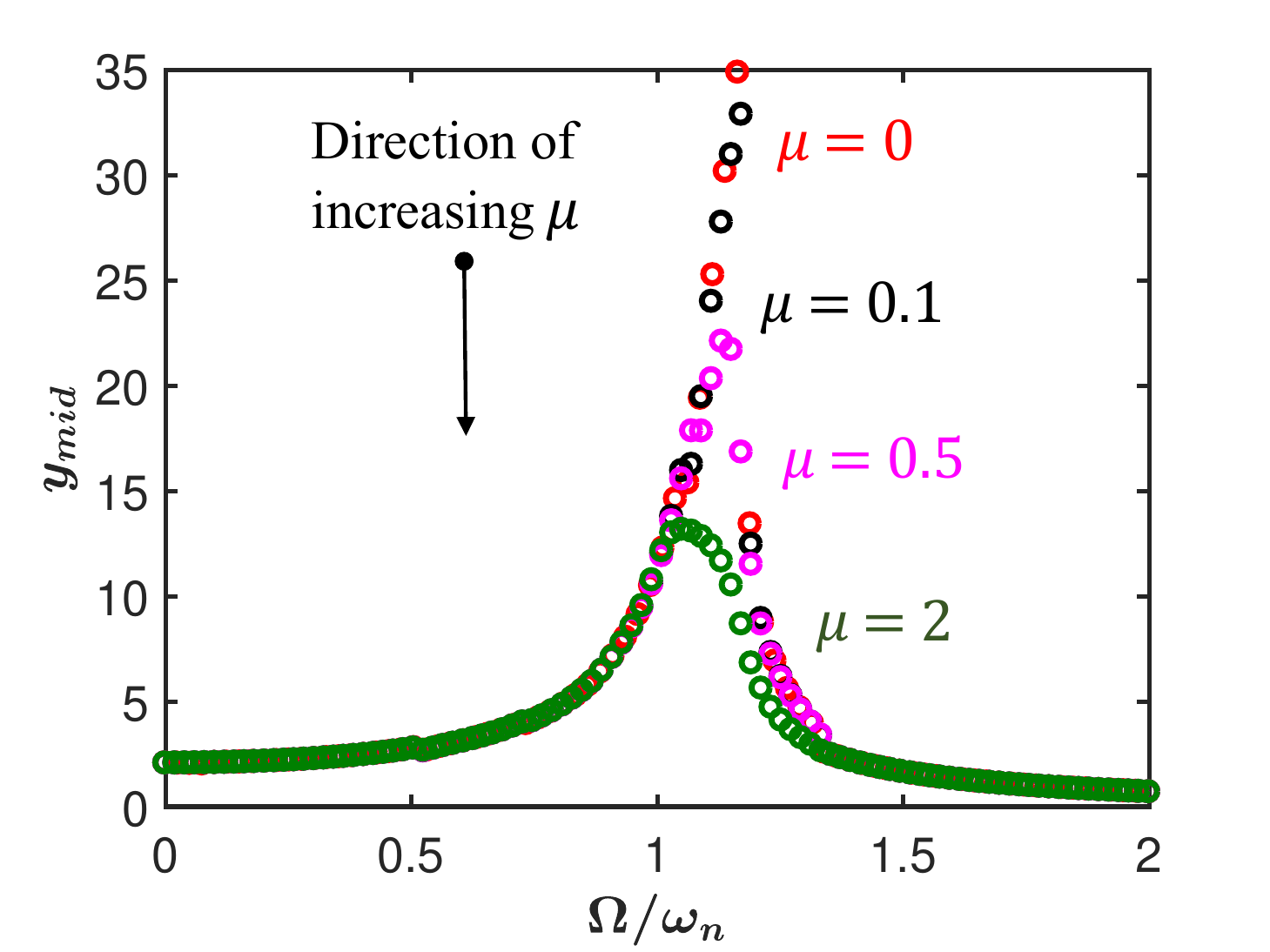}
\label{F5b}}
\caption{(a) Amplitude-frequency behavior at steady state for frictionless biomimetic scale  beam with $\theta_0= 5^\circ$ and various $\eta$. (b) The amplitude-frequency response at steady state for a frictional biomimetic scale beam with $\eta= 5$ and various $\mu$.}
\vspace{-10pt} 
\label{fig2} 
\end{figure}

Next, we probe the forced-vibration behavior of this system by obtaining the amplitude-frequency response. We plot the maximum mid-point deflection of the beam on the convex side with forcing frequency. We first isolate the effect of $\eta$ by neglecting interfacial friction. However, a small artificial viscosity $C=0.5$ (which amounts to a damping ratio of $\approx 0.025$) is used to encourage convergence. Without loss of generality, the forcing amplitude is taken to be $f_0=200$ and initial scale inclination $\theta_0=5^\circ$. The results of this simulation are shown in Fig.~\ref{F5a} in which we plot the amplitude versus the non-dimensional applied frequency(normalized by the natural frequency of a plain beam). In this figure, we get a pronounced backbone curve, similar to a hardening nonlinear oscillator. We observe that increasing $\eta$ lead to a greater flip over behavior indicating a stronger nonlinearity, taking the resonance farther away from that of a plain beam. Next, in Fig.~\ref{F5b} we investigate the effect of friction. We take $\eta=5$ and vary $\mu$. As in the case of free vibration, the role of friction is found to be mainly that of dampener in forced vibration response. The resonant frequency is therefore moved closer to the natural frequency of the substrate with increasing friction.  

In conclusion, although previous static deflection studies had highlighted the dual role of friction as a dissipater as well as stiffness enhancer, the dynamic studies revealed phenomena that are even more interesting. We found that in spite of postulating only dry Coulomb friction, the oscillations resemble viscous damping behavior. The scales provide significant anisotropy of frictional behavior, which can be tailored using interfacial and geometric parameters. We have also found various degrees of sensitivity to initial displacements depending upon scale geometry and friction. Unlike the static case, no appreciable stiffening effect of friction on the dynamics was observed. This was also confirmed in forced vibration studies. Therefore, our study indicates that biomimetic scale beams can be an excellent platform for synthesizing substrates with tailorable damping indices.

\acknowledgments
This work was supported by the United States National Science Foundation’s Civil, Mechanical, and Manufacturing Innovation Grant No. 1825801.

\bibliographystyle{eplbib}
\bibliography{MainFile}

\end{document}